\documentclass[%
reprint,
superscriptaddress,
amsmath,amssymb,
aps,
prb,
]{revtex4-2}
\usepackage{graphicx}
\usepackage{dcolumn}
\usepackage{bm}
\usepackage{chemformula}
\usepackage{hyperref}

\usepackage{tabularx}
\usepackage{amsmath}
\usepackage{float}
\usepackage{booktabs}

\begin{document}
	
\title{Excitations in the ordered and paramagnetic states of honeycomb magnet \ch{Na_2Co_2TeO_6}}
	
\author{Weiliang Yao}
\email{wyao4@utk.edu}
\altaffiliation{Present address: Department of Physics, University of Tennessee, Knoxville, Tennessee 37996, USA}
\affiliation{International Center for Quantum Materials, School of Physics, Peking University, Beijing 100871, China}
\author{Kazuki Iida}
\affiliation{Neutron Science and Technology Center, Comprehensive Research Organization for Science and Society, Tokai, Ibaraki 319-1106, Japan}
\author{Kazuya Kamazawa}
\affiliation{Neutron Science and Technology Center, Comprehensive Research Organization for Science and Society, Tokai, Ibaraki 319-1106, Japan}
\author{Yuan Li}
\email{yuan.li@pku.edu.cn}
\affiliation{International Center for Quantum Materials, School of Physics, Peking University, Beijing 100871, China}
	
\date{\today}%
	
\begin{abstract}
{
\ch{Na_2Co_2TeO_6} is a proposed approximate Kitaev magnet, yet its actual magnetic interactions are elusive due to a lack of knowledge on the full excitation spectrum. Here, using inelastic neutron scattering and single crystals, we determine the system's temperature-dependent magnetic excitations over the entire Brillouin zone. Without committing to specific models, we unveil a distinct signature of the third-nearest-neighbor coupling in the spin waves, which signifies the associated distance as an emerging ``soft link'' in the ordered state. The presence of at least six non-overlapping spin-wave branches is at odds with all models proposed to date. Above the ordering temperature, persisting dynamic correlations can be described by equal-time magnetic structure factors of a hexagonal cluster, which reveal the leading instabilities. Our result sets definitive constraint on theoretical models for \ch{Na_2Co_2TeO_6} and provides new insight for the materialization of the Kitaev model.
}
\end{abstract}
\maketitle

A quantum spin liquid (QSL) is a novel state of matter where localized spins defy formation of long-range order due to frustrated interactions and/or quantum fluctuations \cite{BalentsNature2010,ZhouRMP2017,BroholmScience2020}. The concept has stimulated intense research ever since the original proposal of resonating valence bonds by Anderson \cite{Anderson1973}. In recent years, the spin-1/2 Kitaev honeycomb model has become another booming direction to search for QSLs \cite{Winter2017,Takagi2019,MotomeJPSJ2020,TrebstPR2022}. In this model, spins with bond-dependent Ising interactions (Kitaev interactions) are highly frustrated, and they form QSL ground states along with fractionalized excitations \cite{Kitaev2006}.
	
Materialization of the Kitaev model is illuminated by a mechanism proposed by Jackeli and Khaliullin \cite{JackeliPRL2009} in Mott insulators with strong spin-orbit coupling (SOC). $\alpha$-\ch{RuCl_3} and \ch{Na_2IrO_3} are two representative candidates, where the \ch{Ru^{3+}} and \ch{Ir^{4+}} ions have a low-spin $d^5$ electronic configuration and an atomic ground state of a spin-orbit entangled Kramers doublet \cite{PlumbPRB2014,ChaloupkaPRL2010,SinghPRB2010}. The edge-sharing \ch{RuCl_6} and \ch{IrO_6} octahedra form layered honeycomb lattices, which host nearest-neighbor Kitaev interactions \cite{JackeliPRL2009}. Even though neither system has a QSL ground state under ambient condition, experiments have suggested a major role of Kitaev interactions in the magnetic models \cite{BanerjeeNM2016,BanerjeeScience2017,ChunNP2015,Kim2020,Takagi2019}, whereas the deviation from QSL states is attributed to the presence of additional non-nearest-neighbor-Kitaev terms \cite{ChaloupkaPRL2013,KatukuriNJP2014,RauPRL2014}. Moreover, evidence for a QSL state has been reported in $\alpha$-\ch{RuCl_3} under in-plane magnetic fields \cite{SearsPRB2017,KasaharaNature2018,BanerjeeNPJQM2018,YokoiScience2021,TanakaNP2022}, which have become widely used for the search of QSLs in putative Kitaev magnets with long-range order.
	
Furthering the Jackeli–Khaliullin mechanism, recent theoretical studies indicate that Kitaev interactions can arise between 3$d$ transition-metal ions with a high-spin $d^7$ electronic configuration ($t_{2g}^{5}e_{g}^{2}$) \cite{LiuPRB2018,SanoPRB2018,LiuPRL2020,MotomeJPCM2020,KimJPCM2021_2}. While both \ch{Co^{2+}} and \ch{Ni^{3+}} ions can serve for this purpose \cite{MotomeJPCM2020}, materials studied so far are mostly Co-based, because \ch{Ni^{3+}} is an uncommon oxidation state in solids. Co-based candidate Kitaev magnets include \ch{Na_2Co_2TeO_6} \cite{ViciuJSSC2007,LefrancoisPRB2016,BeraPRB2017,XiaoCGD2019}, $A_3$\ch{Co_2SbO_6} (with $A =$ Li, Na and Ag) \cite{StratanNJC2019,ViciuJSSC2007,WongJSSC2016,YanPRB2019,ZverevaDT2016}, \ch{CoTiO_3} \cite{IshikawaJPSJ1958,YuanPRX2020}, \ch{BaCo_2(AsO_4)_2} \cite{RegnaultPhysicaB1977,ZhongSA2020} and \ch{BaCo_2(PO_4)_2} \cite{NairPRB2018}. Although all of them develop long-range order at low temperatures, the ordering can be suppressed by in-plane fields in \ch{Na_2Co_2TeO_6} \cite{YaoPRB2020,LinNC2021} and \ch{BaCo_2(AsO_4)_2} \cite{ZhongSA2020}, similar to the behavior of $\alpha$-\ch{RuCl_3}. Their thermal transport properties are also similar to $\alpha$-\ch{RuCl_3} \cite{HongPRB2021,ZhongSA2020}.
	
With the promising properties, \ch{Na_2Co_2TeO_6} has recently been intensively studied \cite{YaoPRB2020,SongvilayPRB2020,XPRB2021,LinNC2021,SamarakoonPRB2021,KimJPCM2021,SandersArxiv2021}. A widely recognized goal is to establish the magnetic interaction model with inelastic neutron scattering (INS) \cite{SongvilayPRB2020,LinNC2021,SamarakoonPRB2021,KimJPCM2021,SandersArxiv2021}, yet most of the experiments so far were performed on powder samples and pointed to diversifying sets of parameters. In this work, we report extensive INS data taken on high-quality single crystals, which enable us to map out magnetic excitations over the two-dimensional (2D) Brillouin zone and study their temperature dependence in conjunction with thermodynamics. We find that a third-nearest-neighbor interaction alone provides a highly accurate effective description of the low-energy spin waves, whereas the full spin-wave spectrum \textit{qualitatively rejects} all presently available models. Moreover, the paramagnetic state features persisting short-range magnetic correlations accountable by zigzag-typed magnetization on a hexagonal cluster. These results provide new insights on the closely competing interactions and instabilities in \ch{Na_2Co_2TeO_6}, paving the way to a deeper understanding of Kitaev magnets.
	
\begin{figure}[t!]
	\centering{\includegraphics[clip,width=9cm]{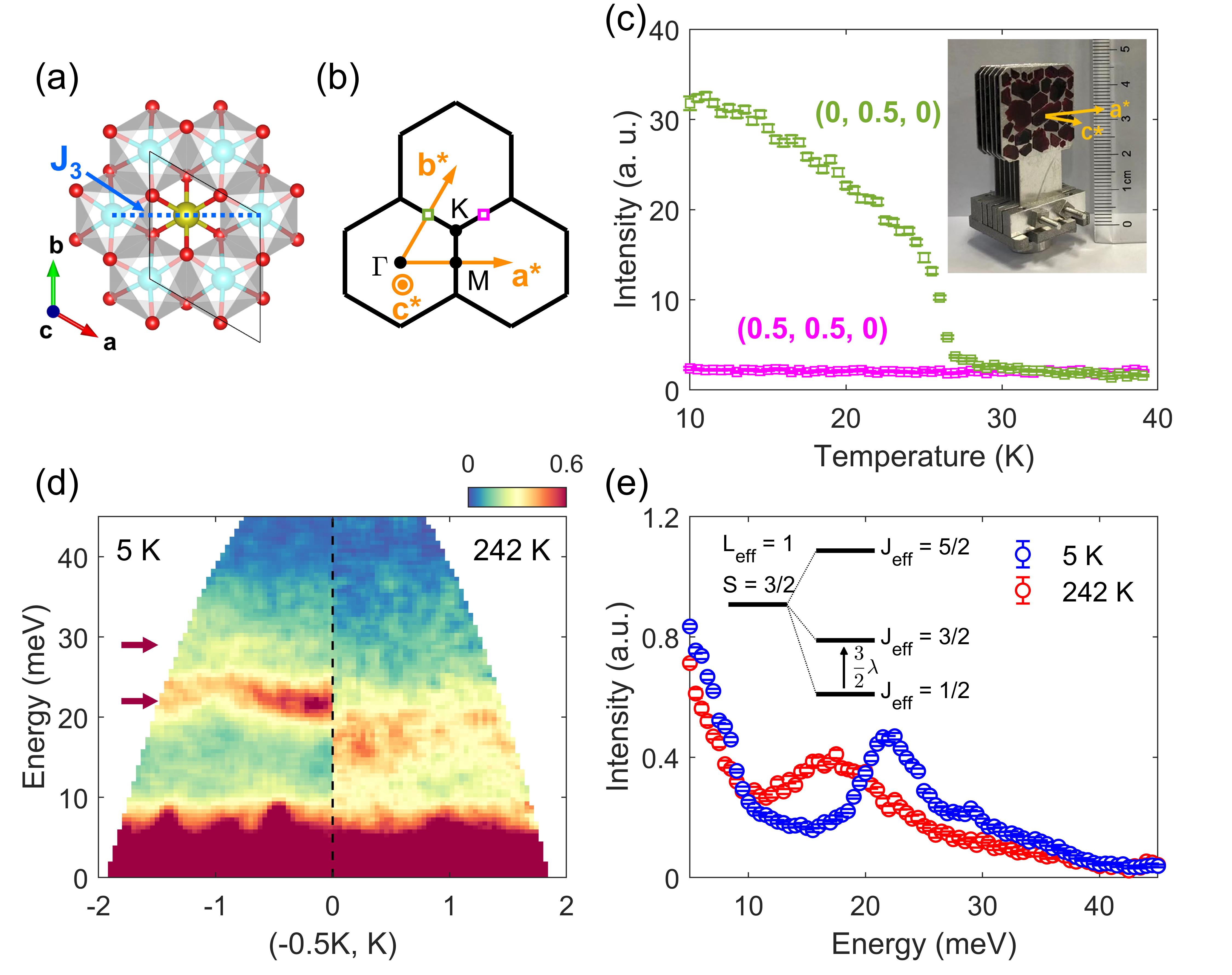}}
	\caption{(a) A honeycomb layer of \ch{Na_2Co_2TeO_6}. Cyan, yellow, and red spheres represent Co, Te and O, respectively. Solid lines indicate a 2D primitive cell. Dotted line connects a pair of third-nearest-neighbor \ch{Co^{2+}} ions. The illustration is produced with VESTA \cite{MommaJAC2011}. (b) 2D reciprocal space and hexagonal Brillouin zones. (c) Diffraction at (0, 0.5, 0) and (0.5, 0.5, 0), measured versus $T$ with $E_\mathrm{i} = 10.0$ meV. The two $\mathbf{Q}$ positions are indicated in (b). Inset shows a photograph of our sample. (d) Crystal-field excitations along $\mathbf{Q}_{\mathrm{2D}}=(-0.5 K, K)$, measured at two temperatures with $E_\mathrm{i} = 52.9$ meV. (e) Energy distribution of intensity, after integrating the (symmetrized) data in (d) over $K \in [-1, 1]$. Inset illustrates the splitting of the 12-fold degenerate atomic $L_{\rm{eff}}$ = 1, $S$ = 3/2 states under the influence of SOC. Arrow indicates the observed excitations in the fully-localized limit.}
	\label{fig1}
\end{figure}

\ch{Na_2Co_2TeO_6} has nearly ideal honeycomb layers of edge-sharing \ch{CoO_6} octahedra [Fig. \ref{fig1}(a)] \cite{ViciuJSSC2007,LefrancoisPRB2016,BeraPRB2017,XiaoCGD2019}. Due to SOC and the octahedral crystal field, the \ch{Co^{2+}} ions in their high-spin configuration are expected to have a pseudospin $J_{\rm{eff}}$ = 1/2 ground state \cite{LiuPRB2018,SanoPRB2018}. Below $T_\mathrm{N}\sim26.5$~K, the system develops long-range three-dimensional antiferromagnetic (AFM) order with a propagation vector (0, 1/2, 0) and its symmetry-related equivalents \cite{BeraPRB2017,XiaoCGD2019,SamarakoonPRB2021}. The precise magnetic structure, however, has some ambiguities: one possibility is a zigzag structure \cite{LefrancoisPRB2016,BeraPRB2017,SamarakoonPRB2021}, which has $C_3$-related domains in a macroscopic sample; another is a ``triple-$\mathbf{q}$'' structure formed by the vector sum of all $C_3$-related zigzag structures \cite{XPRB2021}, which was originally discussed as a field-induced state \cite{JanssenPRL2016}. Difficult to distinguish in most experiments, these two structures are both referred to as zigzag-typed in the present study.

Single crystals of \ch{Na_2Co_2TeO_6} were grown by a modified flux method described in \cite{SM}. About 200 single crystals ($\sim2$ grams in total) were coaligned with reciprocal vectors $\textbf{a}^*$ and $\textbf{c}^*$ horizontal [Fig. \ref{fig1}(b) and (c) inset]. The INS experiment was performed on the 4SEASONS time-of-flight spectrometer at the MLF, J-PARC, Japan \cite{KajimotoJPSJS2011}, using a main incident neutron energy $E_\mathrm{i} = 10.0$ meV and Fermi chopper frequency 150 Hz. Data from additional $E_\mathrm{i}$'s (2.9, 4.1, 6.1, 19.4, and 52.9 meV) were obtained simultaneously \cite{NakamuraJPSJ2009}. Sample-rotation (``4D'') measurements were performed at nine temperatures ($T = 5$, 14, 21, 28, 35, 63, 120, 242, and 290 K). Data were analyzed with Utsusemi \cite{InamuraJPSJ2013}, Horace \cite{EwingsHorace2016} and DAVE \cite{AzuahNIST2009}. All intensities except for those obtained with $E_\mathrm{i} = 52.9$ meV were converted to absolute units \cite{XuRSI2013} using phonon scattering around (3,~0,~0) \cite{SM}. To present excitations in the ($H$, $K$) plane, the normalized data were averaged over the entire covered $L$-range. Spin-wave calculations were performed with SpinW \cite{TothJPCM2015}. Specific heat measurements were performed on a single crystal with a Quantum Design PPMS, where the magnetic specific heat was obtained by subtracting lattice contributions measured on a non-magnetic \ch{Na_2Zn_2TeO_6} reference crystal \cite{YaoPRB2020}.
	
Since variations of ordering temperatures caused by sample imperfection have greatly complicated the interpretation of results in $\alpha$-\ch{RuCl_3} \cite{BanerjeeNM2016,CaoPRB2016}, a pre-check of the magnetic ordering in our \ch{Na_2Co_2TeO_6} crystal array is desired. Figure \ref{fig1}(c) presents the $T$ dependence of a magnetic Bragg peak at (0, 0.5, 0). The observed transition around 26.5 K is consistent with thermodynamically determined $T_\mathrm{N}$ \cite{BeraPRB2017,XiaoCGD2019}, confirming the high homogeneity of our sample. No temperature dependence is found for the intensity at (0.5, 0.5, 0), which rules out the so-called stripe-typed magnetic order \cite{ChoiPRL2012}.
	
Given the relatively weak SOC in 3$d$ transition metals, the pseudospin picture is not necessarily adequate for describing the low-energy physics in Co-based compounds \cite{KimPRB2020,KimJPCM2021_2}. To check this, we inspect the crystal-field excitations of \ch{Na_2Co_2TeO_6}. As presented in Fig. \ref{fig1}(d)-(e), \textit{two} excitation levels can be observed between 20 and 30 meV at 5 K. The more pronounced one around 22 meV has clear dispersion along $(-0.5 K, K)$, and its intensity distribution in the ($H$, $K$) plane can be found in \cite{SM}. Well above $T_\mathrm{N}$, the excitations move to lower energy due to vanishing molecular fields associated with the long-range magnetic order, which can be more clearly seen from the energy distribution plot in Fig.~\ref{fig1}(e). A zeroth-order approximation to these excitations is the process of exciting electrons from $J_{\rm{eff}}$ = 1/2 to $J_{\rm{eff}}$ = 3/2 states \cite{SongvilayPRB2020,KimPRB2020,KimJPCM2021_2}, schematically showed in the inset of Fig.~\ref{fig1}(e). Hence, the persistence of the excitations to far above $T_\mathrm{N}$ supports the validity of the $J_{\rm{eff}}$ = 1/2 picture in \ch{Na_2Co_2TeO_6}. The non-zero dispersion of the 22 meV band, and the presence of a weaker high-energy side-band close to 30 meV at 5 K, are likely due to electron itinerancy and inter-mixing between the $J_{\rm{eff}} = 1/2$ and $3/2$ states \cite{BuyersJPC1971} under additional non-octahedral crystal fields.

\begin{figure}[t!]
\centering{\includegraphics[clip,width=8cm]{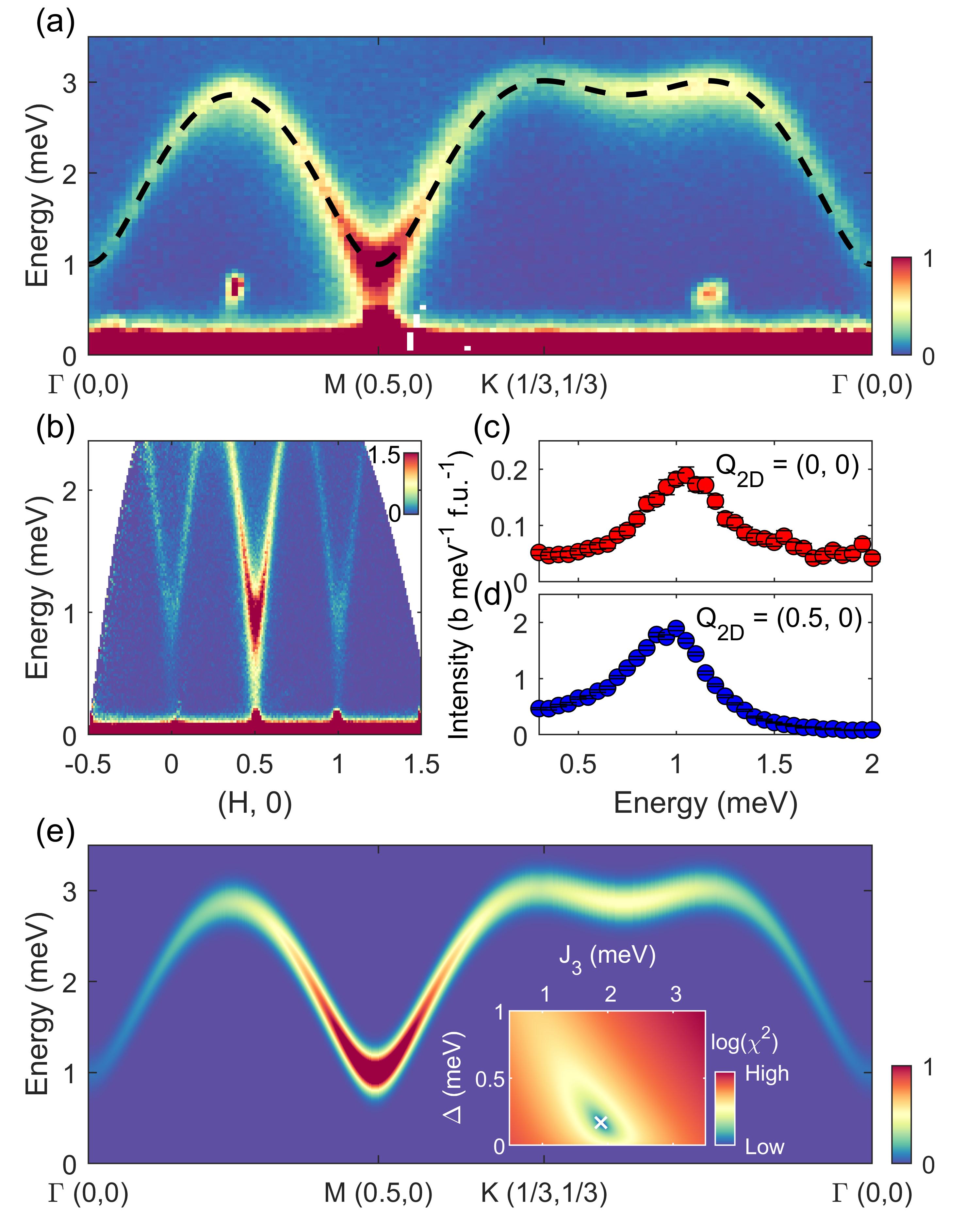}}
\caption{(a) Low-energy spin waves along high-symmetric lines of the Brillouin zone [Fig.~\ref{fig1}(b)], measured with $E_\mathrm{i} = 6.1$ meV. Two singular signals below 1 meV are artifacts (multiple scattering). Dotted line is a fit dispersion, see text. (b) Band bottoms of spin waves along ($H$, 0), measured with $E_\mathrm{i} = 2.9$ meV. (c) and (d) Energy cuts at (0, 0) and (0.5, 0), based on the same data as in (b). Slight difference in the peak-maximum energy is due to resolution effects. (e) Calculated spin waves using the model in Eq.~(\ref{Eq1}) for comparison to (a). Inset shows the goodness of fit ($\chi^2$) versus $J_3$ and $\Delta$. White cross indicates the best-fit parameters used for the main panel.}
\label{fig2}
\end{figure}

\begin{figure}[t!]
\centering{\includegraphics[clip,width=9cm]{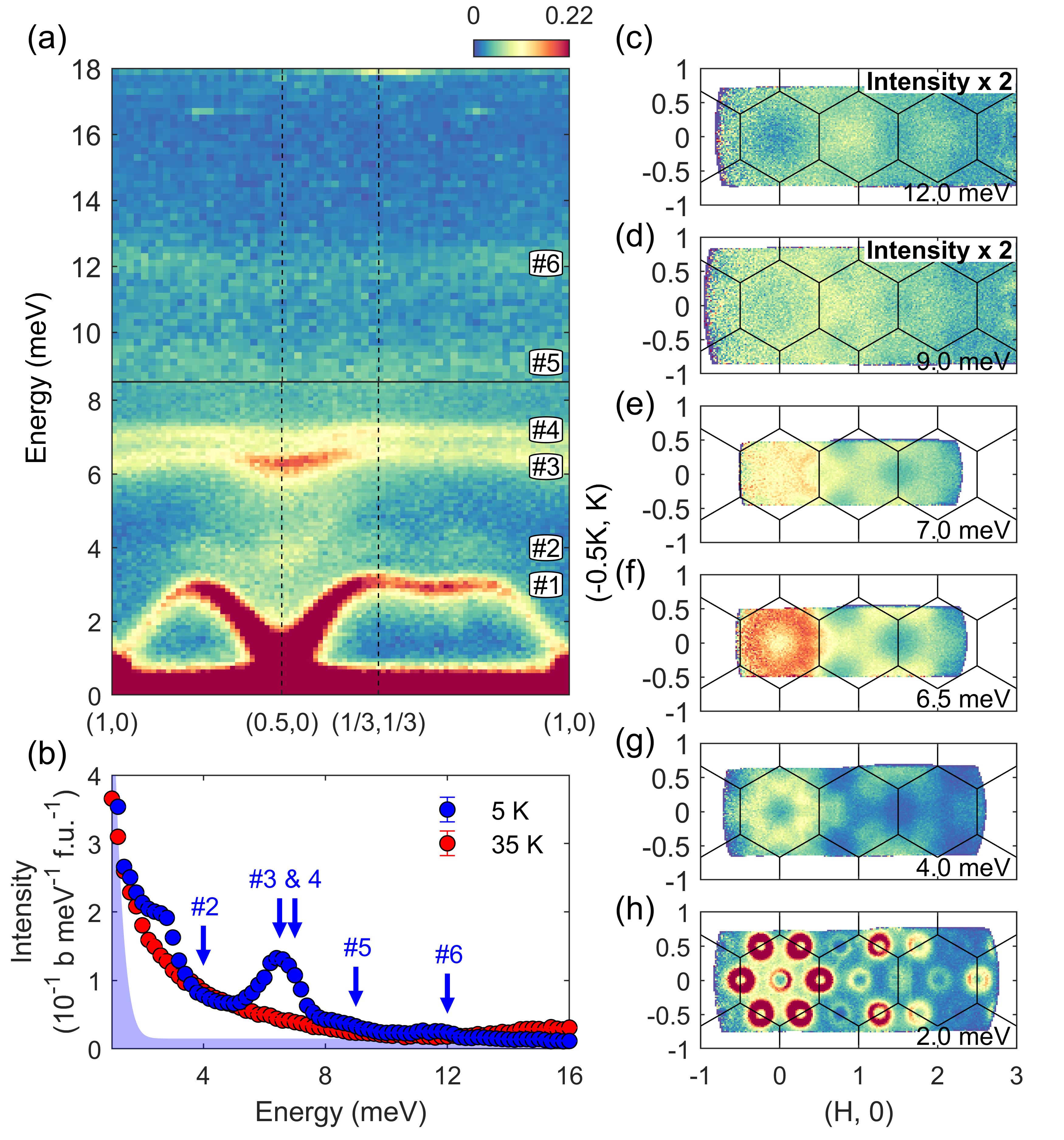}}
\caption{(a) At least six spin-wave branches are observed at $T=5$ K. Data are measured with $E_\mathrm{i} = 10.0$ meV (lower part) and 19.4 meV (upper part). (b) Brillouin-zone averaged intensity versus energy, measured with $E_\mathrm{i} = 19.4$ meV. Shaded area indicates background scattering (excluded from the sum-rule analysis discussed in the text). The slightly increased intensity above 12 meV at 35 K is contributed by the softened crystal field excitations (Fig.~\ref{fig1}). (c)-(h) Spin-wave signals at selected energies.}
\label{fig3}
\end{figure}
	
Next, we turn to magnetic excitations within the $J_{\rm{eff}}$ = 1/2 manifold. Figure \ref{fig2}(a) presents the lowest-energy spin wave branch along high-symmetric lines in the 2D Brillouin zone. This branch reaches its energy bottom ($\sim$1 meV) at the M-point, consistent with previous INS results \cite{SongvilayPRB2020,LinNC2021,SamarakoonPRB2021,KimJPCM2021,SandersArxiv2021}. However, a closer look along ($H$, 0) [Fig.~\ref{fig2}(b)-(d)] indicates that the same energy bottom is also reached at the $\Gamma$-point. According to spin-wave theory, if zigzag order is primarily stabilized by Kitaev interactions, the spin waves are expected to be flat modes near the $\Gamma$-point \cite{ChaloupkaPRL2013,BanerjeeNM2016}, which is clearly different from our observation. The identical dispersion near the M- and the $\Gamma$-points was previously taken as a key support for a triple-$\mathbf{q}$ magnetic structure \cite{XPRB2021}.

Local-moment models usually have prominent nearest-neighbor interactions, as has also been inferred from powder INS data \cite{SongvilayPRB2020,LinNC2021,SamarakoonPRB2021,KimJPCM2021,SandersArxiv2021}. Surprisingly, the lowest-energy spin waves can be adequately described by an effective model with \textit{only} third-nearest-neighbor AFM coupling ($J_3$) and gap-opening anisotropy ($\Delta$) terms:
\begin{equation}
	H = J_3 \sum_{ \langle\langle\langle i,j \rangle\rangle\rangle}\textbf{S}_i\cdot\textbf{S}_j-\Delta\sum_{i}\left(\textbf{S}_i\cdot\hat{n}_i\right)^2.
\label{Eq1}
\end{equation}
The model has N\'{e}el order on each of the four $J_3$-linked (enlarged honeycomb) sub-lattices, and $\hat{n}_i$ denotes the ordered spin direction at site $i$. Using $J_3 = 1.896(9)$ meV and $\Delta = 0.170(6)$ meV, the calculated dispersion [Fig. \ref{fig2}(a)] and dynamic structure factor [Fig. \ref{fig2}(e)] agree very well with our INS data. The inset of Fig.~\ref{fig2}(e) shows a globally optimal parameter set (for detail, see \cite{SM}). We attribute the success of this model to an emerging network of $J_3$ in the AFM ordered state, and make two remarks: (1) The model is compatible with all zigzag-typed structures as they are degenerate ground states. In the limit that the inter-sub-lattice interactions are cancelled in the ordered structure, the low-energy dynamics will be dictated by the effective $J_3$ and $\Delta$. (2) Taking a metaphor to a crystal of organic molecules: the lowest-energy phonons will reflect the weak inter-molecular coupling (\textit{e.g.}, hydrogen bonds and van der Waals forces), rather than the strong intra-molecular coupling (\textit{e.g.}, covalent bonds). Similarly, without knowing the bare exchange interactions, $J_3$ in our model could be an effective coupling that derives from the bare interactions under a frustrated order, which features small magnetic clusters linked by the effective $J_3$.

\begin{figure*}[t!]
\centering{\includegraphics[width=1\textwidth]{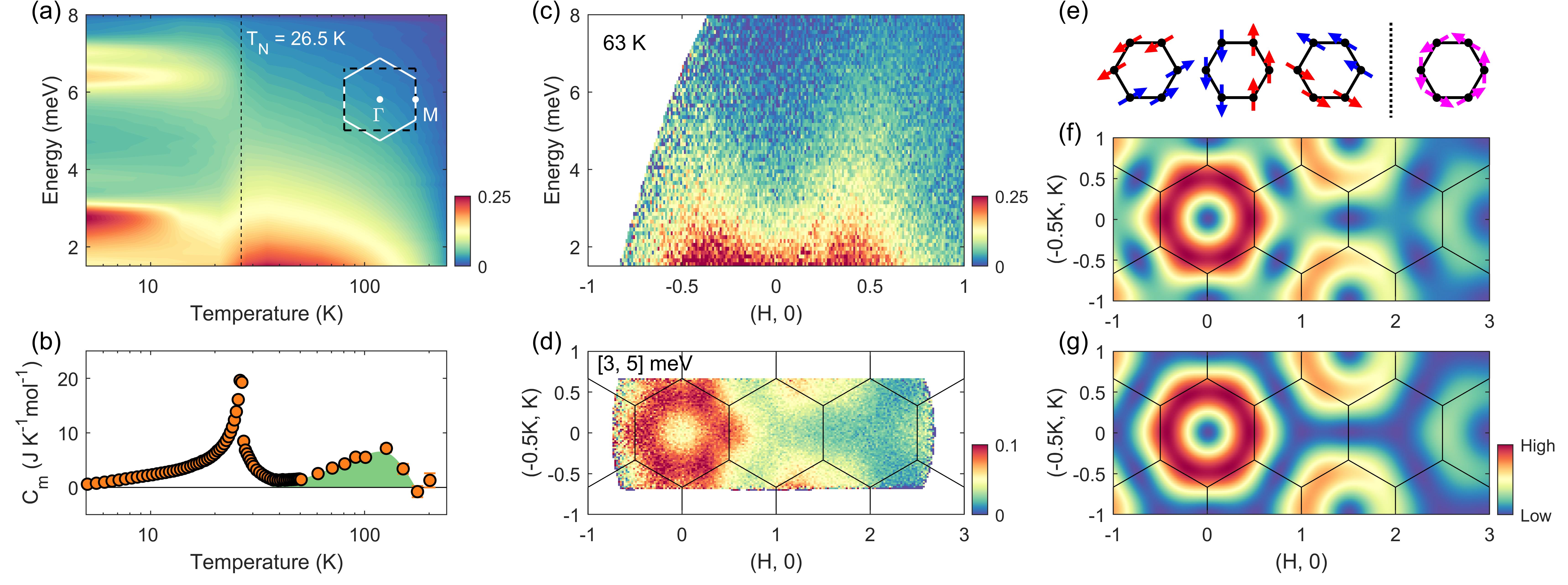}}
\caption{(a) Energy and temperature dependence of intensity averaged over a Brillouin zone (dashed black rectangle in inset), based on data obtained at 8 temperatures with $E_\mathrm{i} = 10.0$ meV, after subtraction against $T=290$ K as background. (b) Magnetic specific heat. Shaded area indicates heat release above $T_\mathrm{N}$. (c) and (d) Paramagnetic fluctuations measured with $E_\mathrm{i} = 10.0$ meV at 63 K. (e) Zigzag spin arrangements on a hexagonal unit related by $C_3$ rotation (left), and their vector superposition forming a tornado-like cluster (right). The calculated structure factors are shown in (f) and (g), respectively.}
\label{fig4}
\end{figure*}

At higher energy up to 12 meV, we observe at least five weakly-dispersing excitation branches [Fig. \ref{fig3}(a)]. We attribute them to additional spin waves, because they completely disappear above $T_\mathrm{N}$ [Fig.~\ref{fig3}(b)] and have a rich variety of dynamic structure factors at 5 K [Fig.~\ref{fig3}(c)-(h)]. The factor of $\sim2$ energy hierarchy compared to the crystal-field excitations provides an estimate of how good the $J_{\rm{eff}} = 1/2$ description is. By applying a sum-rule analysis \cite{XuRSI2013,LorenzanaPRB2005,SM}, we obtain a total spectral weight (from 1 meV to 14 meV) corresponding to $g^2S \approx 7.53$ at 5 K. The inferred $g$-factor (for simplicity, assumed to be a scalar) of $\sim4$ for effective $S=1/2$ is consistent with electron paramagnetic resonance measurements \cite{LinNC2021}. While a complete model for the spin waves is beyond the scope of this study (in part because the ground-state structure is unknown), some key characteristics are noted: (i) The lowest-energy branch carries most of the spectral weight and thus dominates the dynamic correlations. (ii) The next most pronounced branches, $\#$3 and $\#$4 in Fig.~\ref{fig3}(b), have qualitatively similar dispersion (\textit{i.e.}, same energy minimum reached at both the M- and $\Gamma$-points) and $S(\mathbf{Q})$ as the lowest branch [Fig.~\ref{fig3}(a)], which suggests that the effective $J_3$ is also important for them. (iii) The number of spin-wave branches sets a lower bound on the number of spins in the magnetic primitive cell. The branches have no overlap, which is distinct from other honeycomb magnets with branch crossings \cite{ChoiPRL2012,ChenPRX2018,YuanPRX2020,GaoPRB2021}. This further hints at the existence of magnetic clusters \cite{FurrerRMP2013} in the ordered state.
	
We have compared to spin waves calculated from published models (see \cite{SM} for the actual comparisons), and found all of them to be qualitatively inconsistent with our INS data, especially concerning characteristic (iii) above. Once averaged over sample orientations (Fig. S5 in \cite{SM}), our data are fully consistent with powder INS spectra \cite{SongvilayPRB2020,LinNC2021,SamarakoonPRB2021,KimJPCM2021,SandersArxiv2021}, including having a concave $E$-$Q$ envelope shape at small $Q$ near the M-point, which has been taken as a key indication for zigzag order in $\alpha$-\ch{RuCl_3} and \ch{Na_2IrO_3} \cite{BanerjeeNM2016,ChoiPRL2012,Takagi2019}. We believe that further theoretical work is needed to coherently account for the elusive magnetic ground state, the multiple thermal \cite{YaoPRB2020,XPRB2021} and field-induced transitions \cite{YaoPRB2020,LinNC2021}, and the spin waves in \ch{Na_2Co_2TeO_6}. Our extensive INS data provide a solid ground for such explorations.

The physical essence of our effective $J_3$ may be important. The inclusion of $J_3$ on the honeycomb lattice is known to produce rich competing phases in models both with \cite{KatukuriNJP2014,RauPRL2014,SizyukPRB2014,WinterPRB2016,KimchiPRB2011} and without \cite{FouetEPJB2001,MessioPRB2011} anisotropic (\textit{e.g.}, Kitaev) terms. In particular, a classical-energy degeneracy between collinear and non-collinear zigzag-typed states is found in the Heisenberg models \cite{MessioPRB2011}. From a structural point of view, the Co hexagons in \ch{Na_2Co_2TeO_6} are centered around Te atoms, whose spatially-extended $d$ orbitals may promote electron itinerancy and further-neighbor coupling. Even in the cases of $\alpha$-\ch{RuCl_3} and \ch{Na_2IrO_3}, which have no or small-ionic-radius atoms at the hexagon centers, the role of itinerancy \cite{MazinPRL2012,FoyevtsovaPRB2013} and further-neighbor coupling \cite{JanssenPRB2017,MaksimovPRR2020,LaurellNPJQM2020} is being actively discussed in recent years.
	
We last discuss magnetic correlations in the paramagnetic state. They manifest themselves in the INS spectra as an energy down-flow of spin-wave signals from the ordered state [Fig. \ref{fig4}(a)]. The persistence of finite-energy dynamics to far above $T_\mathrm{N}$ is in line with the presence of appreciable magnetic specific heat above $T_\mathrm{N}$ [Fig. \ref{fig4}(b)]. These behaviors again closely resemble $\alpha$-\ch{RuCl_3}, where interpretations have been made around thermodynamics of Majorana fermions \cite{DoNP2017,MotomeJPSJ2020}. We refrain from making related speculations because the microscopic model is unclear at present. Figure \ref{fig4}(c)-(d) shows that the paramagnetic fluctuations are weakly structured in energy, but strongly structured in $\mathbf{Q}$: intensities are concentrated around the M-points, indicative of instability towards the ordering at low $T$. After a widely used method for analyzing frustrated magnets \cite{LeeNature2002,TomiyasuPRL2008,TomiyasuPRB2011,TomiyasuPRB2011_2,JanasPRL2021}, we model the $\mathbf{Q}$ dependence with equal-time spin correlations, by considering scattering interference from a hexagonal unit:
	\begin{equation}
		I(\textbf{Q}) = f^2(\textbf{Q}) \sum_{m,n}e^{i\textbf{Q}\cdot(\textbf{r}_m-\textbf{r}_n)}\sum_{\alpha,\beta}(\delta_{\alpha,\beta}-\frac{Q_{\alpha}Q_{\beta}}{Q^2})\langle S_m^{\alpha}S_n^{\beta} \rangle,
	\end{equation}
where $f(\textbf{Q})$ is the magnetic form factor of \ch{Co^{2+}}, $S_m^{\alpha}$ and $S_n^{\beta}$ are spin components $\alpha$ and $\beta$ at sites $\textbf{r}_m$ and $\textbf{r}_n$, respectively, with $m,\,n \in \{1,\,\ldots,\,6\}$, $\alpha,\,\beta \in \{x,\,y,\,z\}$ and $\delta_{\alpha,\beta}-Q_{\alpha}Q_{\beta}/Q^2$ being a projection factor for unpolarized neutron scattering. $\langle \cdots \rangle$ assumes a 4$\pi$ (global) rotational average of all six spins in the paramagnetic state. For two zigzag-typed arrangements depicted in Fig. \ref{fig4}(e), the above formula can be further simplified as
	\begin{equation}
		I(\textbf{Q}) = \frac{3}{4} f^2(\textbf{Q}) \langle|\sum_{m = 1\dotsb 6} S  e^{i(\phi_m + \textbf{Q} \cdot \textbf{r}_m)}|^2\rangle_{\rm{eq}},
	\end{equation}
where $S$ is the spin size and $\phi_m$ the angle in the honeycomb plane at site $m$, and $\langle \cdots \rangle_{\rm{eq}}$ averages over symmetry equivalents [on the left of Fig. \ref{fig4}(e)].

Satisfactory descriptions of the measurement data are obtained [Fig. \ref{fig4}(g) and (h)] by using both the collinear and non-collinear zigzag-typed clusters. Simulations of other spin arrangements on a hexagon can be found in \cite{SM}. We therefore conclude that the paramagnetic fluctuations are adequately described within one hexagonal unit, and that they are essentially zigzag-typed AFM fragments. A common characteristic of the two arrangements in Fig. \ref{fig4}(e) is that the (presumably AFM) $J_3$ coupling always connects opposite spins, reminding us of $J_3$'s fingerprint on the most pronounced spin waves in the ordered state. Last but not the least, the tornado-like arrangement in Fig. \ref{fig4}(e) can be understood as a non-zero local expected value of the hexagon-flux operator $W_\mathrm{p}$ \cite{Kitaev2006,Takagi2019}. Since $W_\mathrm{p}$ is a local $Z_2$ conserved quantity of the Kitaev model, the paramagnetic fluctuations might have a deep implication on the QSL physics.
	
In conclusion, we have successfully mapped out the magnetic excitations in \ch{Na_2Co_2TeO_6} single crystals. Low-energy dynamics in both the ordered and the thermally disordered states show a strong indication of magnetic coupling between third-nearest neighbors. While the results do not necessarily mean that $J_3$ is a leading interaction, they do suggest the emergence of magnetic clusters featuring the third-nearest distance. Since \ch{Na_2Co_2TeO_6} shares important thermodynamic and spectroscopic characteristics with previous Kitaev-like magnets, we expect our result to stimulate new thinking of Kitaev materials in general, especially in conjunction with structural properties and electron itinerancy.
	
We with to thank Cristian Batista, Wenjie Chen, V. Ovidiu Garlea, Christian Hess, Xiaochen Hong, Lukas Janssen, Chaebin Kim, Wilhelm G. F. Kr\"{u}ger, Ke Liu, Zhengxin Liu, Je-Geun Park, Fa Wang, and Jiucai Wang for discussions. Work at Peking University was supported by the National Basic Research Program of China (Grant No. 2018YFA0305602) and the NSF of China (Grant Nos. 12061131004, 11874069, and 11888101). The INS experiment was performed at the MLF, J-PARC, Japan, under a user program (Proposal No. 2019B0062).

\nocite{apsrev42Control}
\bibliographystyle{apsrev4-2}
	
\bibliography{ref_ins}

\pagebreak
\pagebreak

\widetext
\begin{center}
\textbf{\large Supplemental Material for ``Excitations in the ordered and paramagnetic states of honeycomb magnet \ch{Na_2Co_2TeO_6}''}
\end{center}
\setcounter{equation}{0}
\setcounter{figure}{0}
\setcounter{table}{0}
\setcounter{page}{1}
\makeatletter
\renewcommand{\theequation}{S\arabic{equation}}
\renewcommand{\thefigure}{S\arabic{figure}}

\section{Single Crystal Growth Method}
	Single crystals of \ch{Na_2Co_2TeO_6} were prepared with a flux method. Starting materials of \ch{Na_2CO_3}, \ch{Co_3O_4} and \ch{TeO_2} were grounded thoroughly with a mole ratio of 15.4 : 5.2 : 21.4 and were put into an alumina crucible with a wall-thickness of $\sim3$ mm. The excess \ch{TeO_2} was served as a flux. The final mixture occupied about half volume of the crucible. The crucible was capped with an alumina plate and then put into a box furnace. To avoid overheating, the crucible was padded with another alumina plate (with thickness of $\sim3$ mm). The furnace was heated up to 1050 $^{\circ}$C in 4 hours and maintained at 1050 $^{\circ}$C for 48 hours. Then it was cooled with 6.5 $^{\circ}$C per hour before being turned off at 600 $^{\circ}$C. After the reaction, the alumina crucible was smashed and red single crystals can be selected out of bluish violet residue. The single crystals were further washed with a NaOH solution. Basic characterizations can be found in Ref.~\cite{YaoPRB2020}.
	
	\section{Additional crystal field excitation data}
	
\begin{figure}[t!]
		\centering{\includegraphics[clip,width=9cm]{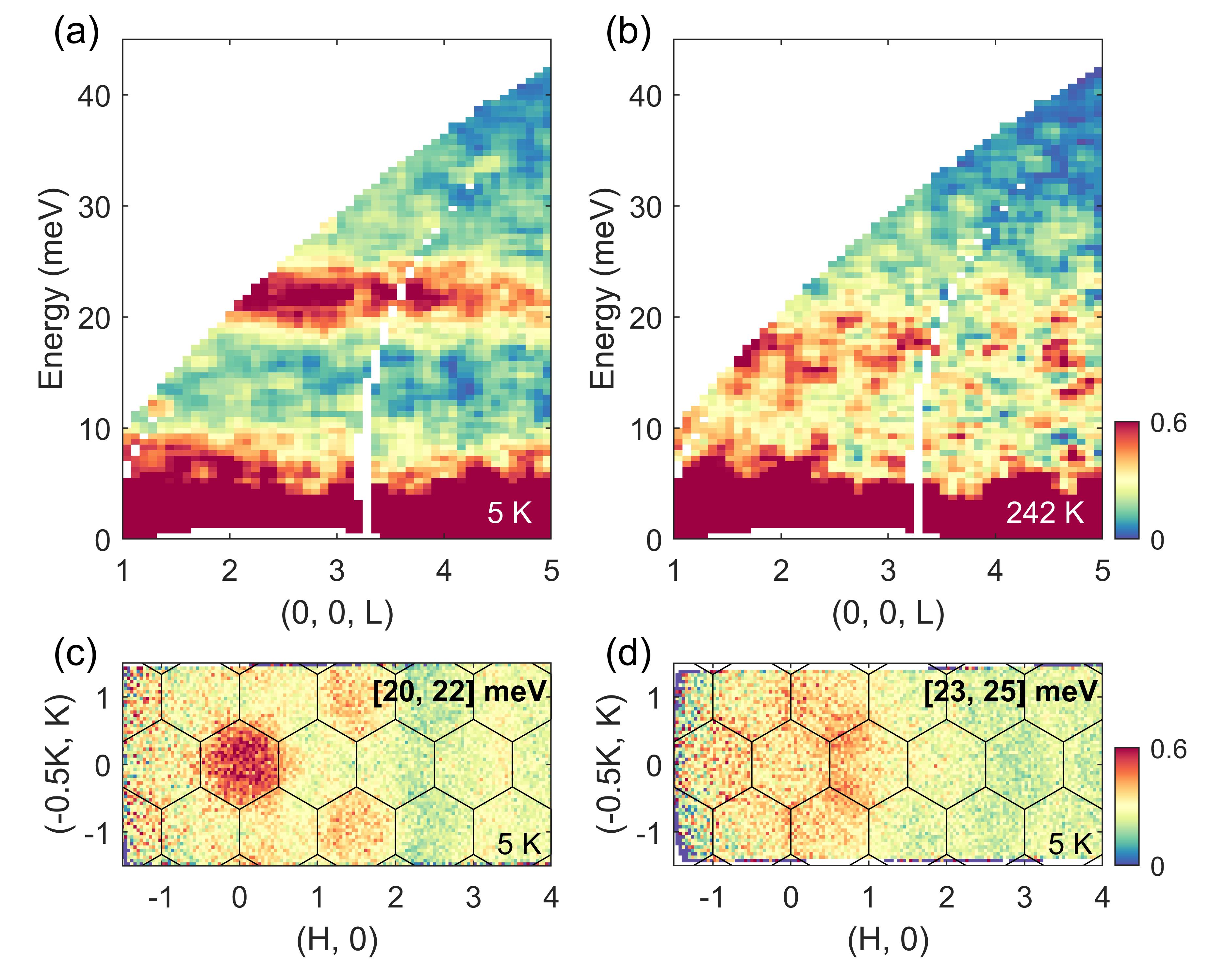}}
		\caption{(a) and (b) Crystal field excitations along (0, 0, $L$) measured with $E_\mathrm{i}$ = 52.9 meV at 5 K and 242 K. (c) and (d) Constant energy cuts in the ($H$, $K$) plane for crystal field excitations around 21 meV and 24 meV at 5 K.}
		\label{figs1}
	\end{figure}

	The crystal field excitation does not show dispersion along (0, 0, $L$) [Fig. \ref{figs1}(a) and (b)]. To present the excitation in the ($H$, $K$) plane, we integrated all measured $L$-range. The dispersive feature of the crystal field excitation can be further seen from the intensity distribution in the ($H$, $K$) plane, as presented in Fig. \ref{figs1}(c) and (d). The intensity concentrating around Brillouin zone center at $\sim21$ meV gradually moves to zone boundaries at $\sim24$ meV.

	\section{Absolute intensity normalization and sum rule}
	
	In the presented data, we normalized the intensities of five $E_\mathrm{i}$'s (2.9, 4.1, 6.1, 10.0 and 19.4 meV) to absolute units according to the phonon around (3, 0, 0). The data of $E_\mathrm{i} = 52.9$ meV were not treated as the low energy part (below 20 meV) is too hard to discern clear magnetic features.
	
	\begin{figure}[t!]
		\centering{\includegraphics[clip,width=10cm]{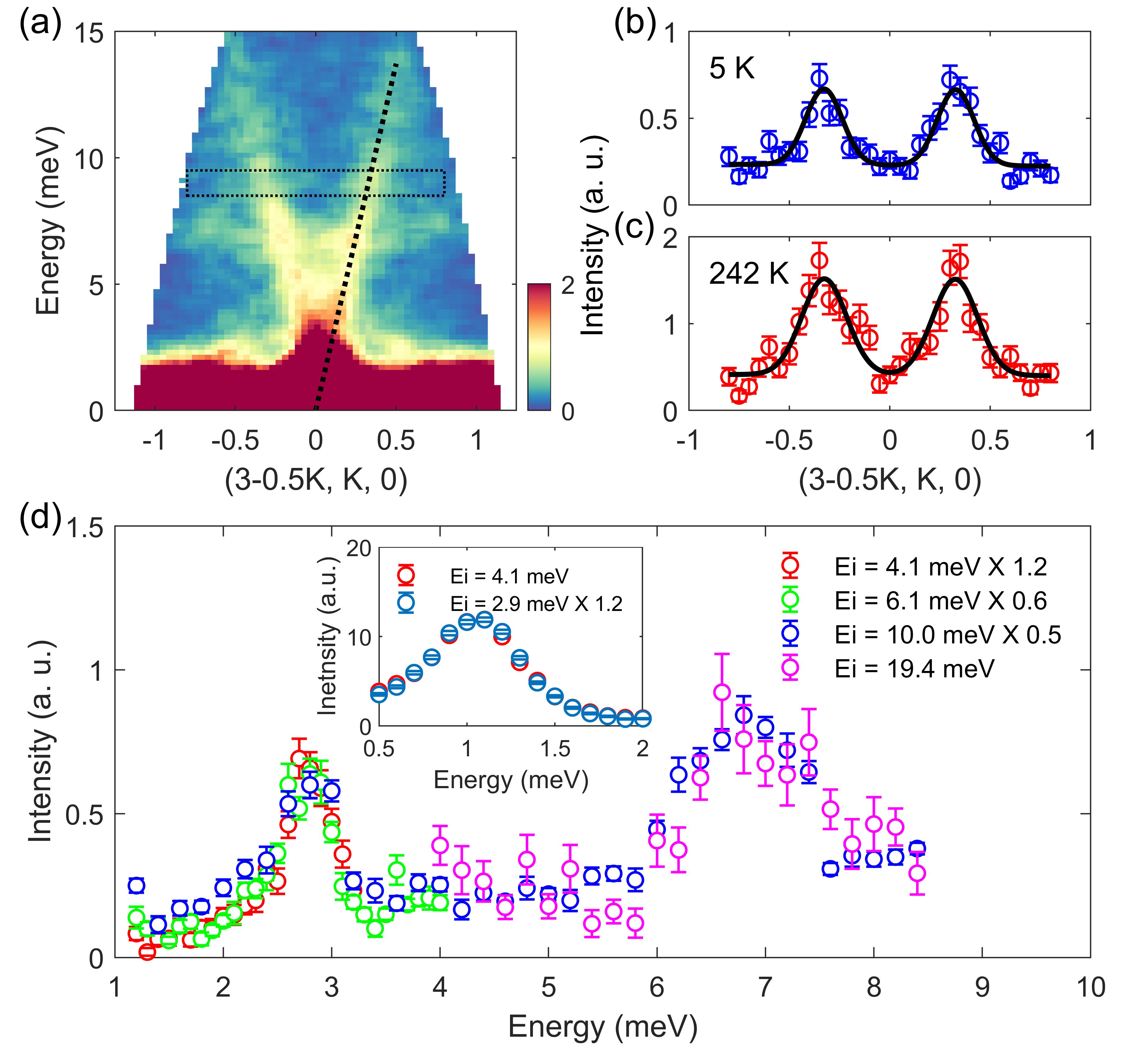}}
		\caption{(a) Acoustic phonon emerging from the Brillouin zone center (3, 0, 0) at 5 K. The data were measured with $E_\mathrm{i}$ = 19.4 meV. The dashed line indicates the phonon dispersion. (b) and (c) Constant energy cuts around 9 meV (dashed rectangle in (a)) at 5 K and 242 K. The solid curves are double-gaussian fits for phonon peaks. (d) Constant momentum cuts around (1.25, 0, 0) with four $E_\mathrm{i}$'s at 5 K. The intensities are normalized according to the data of $E_\mathrm{i}$ = 19.4 meV. Inset shows constant momentum cuts around (0.5, 0, 0) with $E_\mathrm{i}$ = 2.9 meV and $E_\mathrm{i}$ = 4.1 meV at 5 K. The intensities are normalized according to the data of $E_\mathrm{i}$ = 4.1 meV.}
		\label{figs2}
	\end{figure}
	
	With $E_\mathrm{i}$ = 19.4 meV, a branch of acoustic phonon can be observed [Fig. \ref{figs2}(a)]. For neutron scattering, the momentum integrated phonon scattering intensity can be written as \cite{XuRSI2013}
	\begin{equation}
		\int{I(\textbf{Q},\omega)\mathrm{d}\textbf{q}}=\frac{1}{\mathrm{d}\omega/\mathrm{d}q}\frac{n(\omega, T)+1}{\hbar\omega(q)}\frac{(\hbar \textbf{Q})^2}{2m}\frac{m}{M}cos^2(\beta)|F_N(\textbf{G})|^2 e^{-2W} NR_0,
	\end{equation}
	where $\mathrm{d}\omega/\mathrm{d}q$ is the phonon velocity, $n(\omega, T)$ is the Bose factor, $\hbar\omega(q)$ is the phonon energy, $\textbf{Q}$ is the total momentum transfer of the phonon, $\textbf{G}$ is the Brillouin zone center where the phonon locates, $\textbf{q}$ is the momentum transfer relative to $\textbf{G}$, $m$ and $M$ are the masses of a neutron and atoms in one unit cell, respectively, $\beta$ is the phonon polarization angle, $F_N(\textbf{G})$ is the structrue factor of a unit cell, $e^{-2W}$ is Debye-Waller factor and $NR_0$ contains information about sample and instrument (number of unit cells and instrument resolution).
	Constant energy cut for the phonon around 9 meV at 5 K is presented in Fig. \ref{figs2}(b). By integrating over the peak intensity, we can find the left side of (S1) and obtain
	\begin{equation}
		NR_0 = 5.37\,{\rm meV} \cdot {\rm b^{-1}},
	\end{equation}
	through which the absolute intensity for the data of $E_\mathrm{i}$ = 19.4 meV can be obtained. We note the integrated phonon intensity ratio between 242 K and 5 K is 3.10 [Fig. \ref{figs2}(b) and (c)], which is close to the Bose factor ratio of 2.86. Intensities of other $E_\mathrm{i}$s' data can be scaled by factors obtained from constant momentum cuts through low-energy spin waves [Fig. \ref{figs2}(d)].
	
	To apply the sum rule for magnetic neutron scattering, we first note the normalized intensity $\tilde{I} (\textbf{Q},\omega)$ is related to the magnetic dynamic structure factor $\textbf{S}(\textbf{Q}, \omega)$ as \cite{XuRSI2013,LorenzanaPRB2005}
	\begin{equation}
		\tilde{I} (\textbf{Q},\omega) = 0.07266 \,({\rm b}) \, g^2f^2(Q)e^{-2W}\sum_{\alpha,\beta}{\left(\delta_{\alpha,\beta}-\hat{Q}_{\alpha}\hat{Q}_{\beta}\right)S^{\alpha \beta}(\textbf{Q},\omega)},
	\end{equation}
	where $g$ is the Land\'e $g$-factor, $f(Q)$ is the magnetic form factor and the summation is taken over $x$-, $y$- and $z$-components of the spin. The sum rule can be expressed as \cite{XuRSI2013,LorenzanaPRB2005}
	\begin{equation}
		\frac{\sum_{\alpha}\int \mathrm{d}\omega\int_{\mathrm{BZ}}\mathrm{d}\textbf{Q}S^{\alpha \alpha}(\textbf{Q},\omega)}{\int_{\mathrm{BZ}}\mathrm{d}\textbf{Q}}=S(S+1).
	\end{equation}
	We assume the quantized axis of the spin is along $z$ (in spin space) and only the transverse part [$S^{xx} (\textbf{Q},\omega)$ and $S^{yy} (\textbf{Q},\omega)$] are prominent. The sum rule for the spin wave is
	\begin{equation}
		\frac{\int \mathrm{d}\omega\int_{\mathrm{BZ}}\mathrm{d}\textbf{Q}\left[S^{xx}(\textbf{Q},\omega)+S^{yy}(\textbf{Q},\omega)\right]}{\int_{\mathrm{BZ}}\mathrm{d}\textbf{Q}}=S.
	\end{equation}
	Further considering ``domain average'' for the magnetic dynamic structure factor, the normalized spin wave intensity can be written as \cite{LorenzanaPRB2005}
	\begin{equation}
		\tilde{I}_{\mathrm{sw}} (\textbf{Q},\omega) = 0.07266 \,({\rm b}) \, g^2f^2(Q)e^{-2W} \frac{2}{3} \left[S^{xx}(\textbf{Q},\omega)+S^{yy}(\textbf{Q},\omega)\right].
	\end{equation}
	Combining (S5) and (S6), the sum rule can be applied to the normalized spin wave intensity
	\begin{equation}
		\frac{20.64\,({\rm b^{-1}})\int \mathrm{d}\omega\int_{\mathrm{BZ}}\mathrm{d}\textbf{Q}\tilde{I}_{\mathrm{sw}} (\textbf{Q},\omega)}{\int_{\mathrm{BZ}}\mathrm{d}\textbf{Q}}=g^2f^2(Q)e^{-2W}S.
	\end{equation}
	
	Magnetic scattering intensities presented in the main text are obtained by substrating the data at 290 K with the following formula
	\begin{equation}
		I_{\mathrm{mag}}(\textbf{Q}, \omega, T) = I(\textbf{Q}, \omega, T) - \frac{1+n(\omega, T)}{1+n(\omega, 290 \, {\rm K})}I(\textbf{Q}, \omega, 290 \, {\rm K}),
	\end{equation}
	where $I(\textbf{Q}, \omega, T)$ is the measured intensity at momentum $\textbf{Q}$, energy $\omega$ and temperature $T$.
	
	\section{Details on the fit and calculation for the lowest-energy spin wave branch}
	
\begin{figure}[b!]
		\centering{\includegraphics[clip,width=8cm]{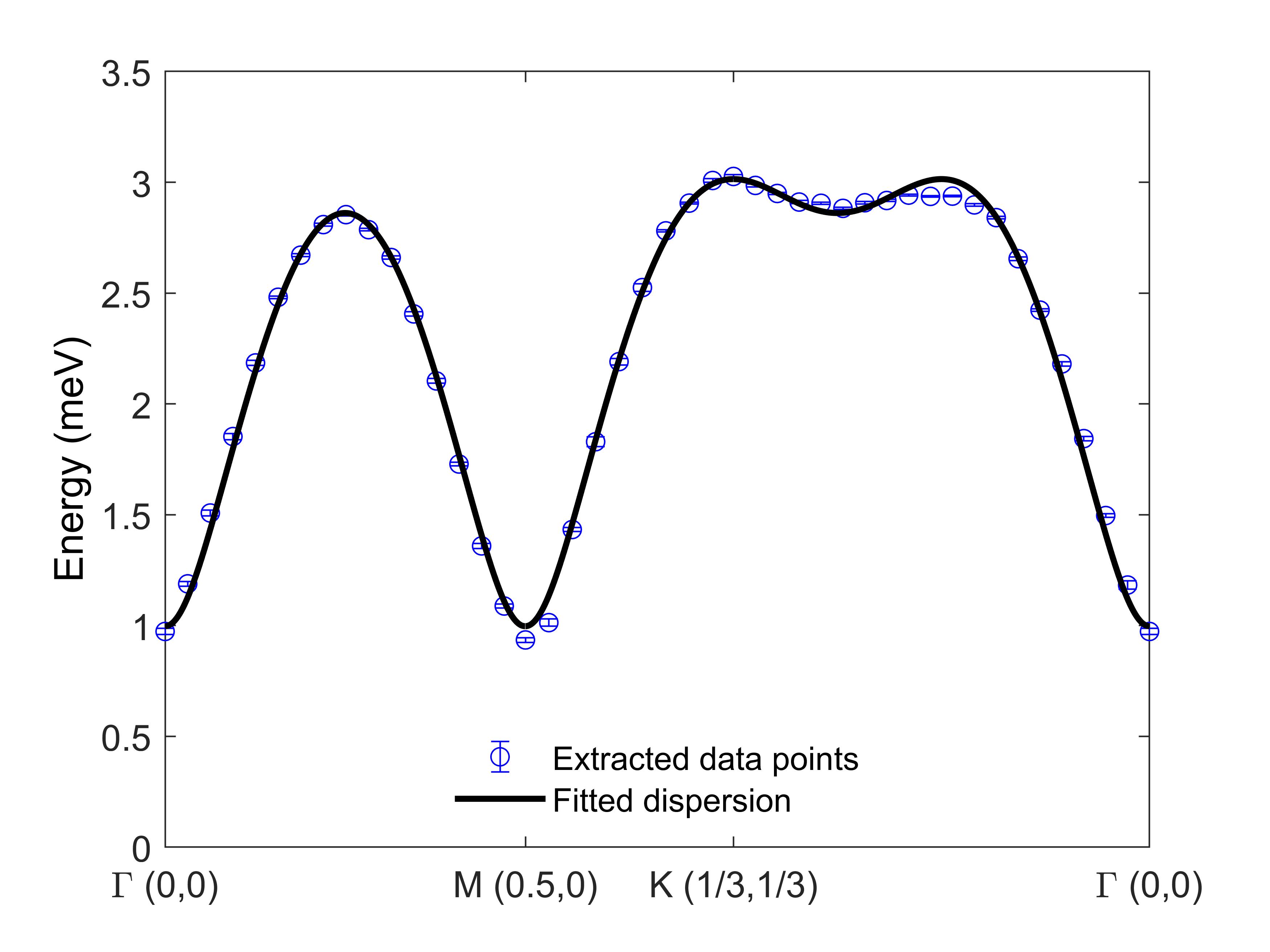}}
		\caption{Fitted data points for the lowest-energy spin wave branch along the trajectory showed in the main text. The solid curve is the calculated dispersion with the optimal parameters.}
		\label{figs3}
	\end{figure}

	The data used in the fit for the spin wave dispersion are presented in Fig. \ref{figs3}, which were obtained by making constant momentum cuts and then making gaussian fits to extract the peak center. The goodness of fit ($\chi^2$) is defined as
	\begin{equation}
		\chi^2=\sum_{\textbf{k}}\frac{\left[\omega_{\mathrm{obs}}(\textbf{k})-\omega(\textbf{k})\right]^2}{\omega(\textbf{k})},
	\end{equation}
	where $\omega_{\mathrm{obs}}(\textbf{k})$ and $\omega(\textbf{k})$ are observed and calculated spin wave energies at momentum transfer $\textbf{k}$, the summation is taken over the sampled $\textbf{k}$ positions (blue dots in Fig. \ref{figs3}).

	The energy eigen values of the Hamiltonian (1) in the main text can be solved out as \cite{Fazekas1999}
	\begin{equation}
		\omega(\textbf{k})=zJ_3S\sqrt{\left(1+\frac{2\Delta}{zJ_3}\right)^2-|\gamma_k|^2},
	\end{equation}
	with
	\begin{equation}
		\gamma_k=\frac{1}{z}\sum_{\bm{\delta}}e^{i\textbf{k}\cdot\bm{\delta}},
	\end{equation}
	where $z$ = 3 (three third-nearest neighbors), $S$ = 1/2  (pseudo-spin quantum number of \ch{Co^{2+}} ions) and the summation is taken over all three third-nearest-neighbor vectors (\bm{$\delta$}).
	
	\section{Calculated spin waves for selective models}
	
	Plenty of microscopic models of \ch{Na_2Co_2TeO_6} have been proposed by fitting the powder INS spectrum \cite{SongvilayPRB2020,LinNC2021,SamarakoonPRB2021,KimJPCM2021,SandersArxiv2021}. All of them are based on the $H$-$K$-$\Gamma$ model with the Hamiltonian written as
	\begin{equation}
		H = \sum_{n = 1,2,3}J_n\sum_{i,j}\textbf{S}_i\cdot\textbf{S}_j+\sum_{i,j}KS_i^{\gamma}S_j^{\gamma}+\sum_{i,j}\Gamma(S_i^{\alpha}S_j^{\beta}+S_i^{\beta}S_j^{\alpha})+\sum_{i,j}\Gamma^{\prime}(S_i^{\alpha}S_j^{\gamma}+S_i^{\gamma}S_j^{\alpha}+S_i^{\beta}S_j^{\gamma}+S_i^{\gamma}S_j^{\beta}),
	\end{equation}
	where $J_n$ with $n$ = 1, 2, 3 are Heisenberg interactions for first-, second- and third-nearest neighbors, $K$ is the Kitaev interaction, $\Gamma$ and $\Gamma^{\prime}$ are bond-dependent off-diagonal interactions, $\alpha$, $\beta$, $\gamma$ denotes the three types of first-nearest-neighbor bonds with \{$\alpha$, $\beta$, $\gamma$\} = \{y, z, x\}, \{z, x, y\}, \{x, y, z\} for X, Y and Z bonds respectively.
	
\begin{figure}[h]
		\centering{\includegraphics[clip,width=14cm]{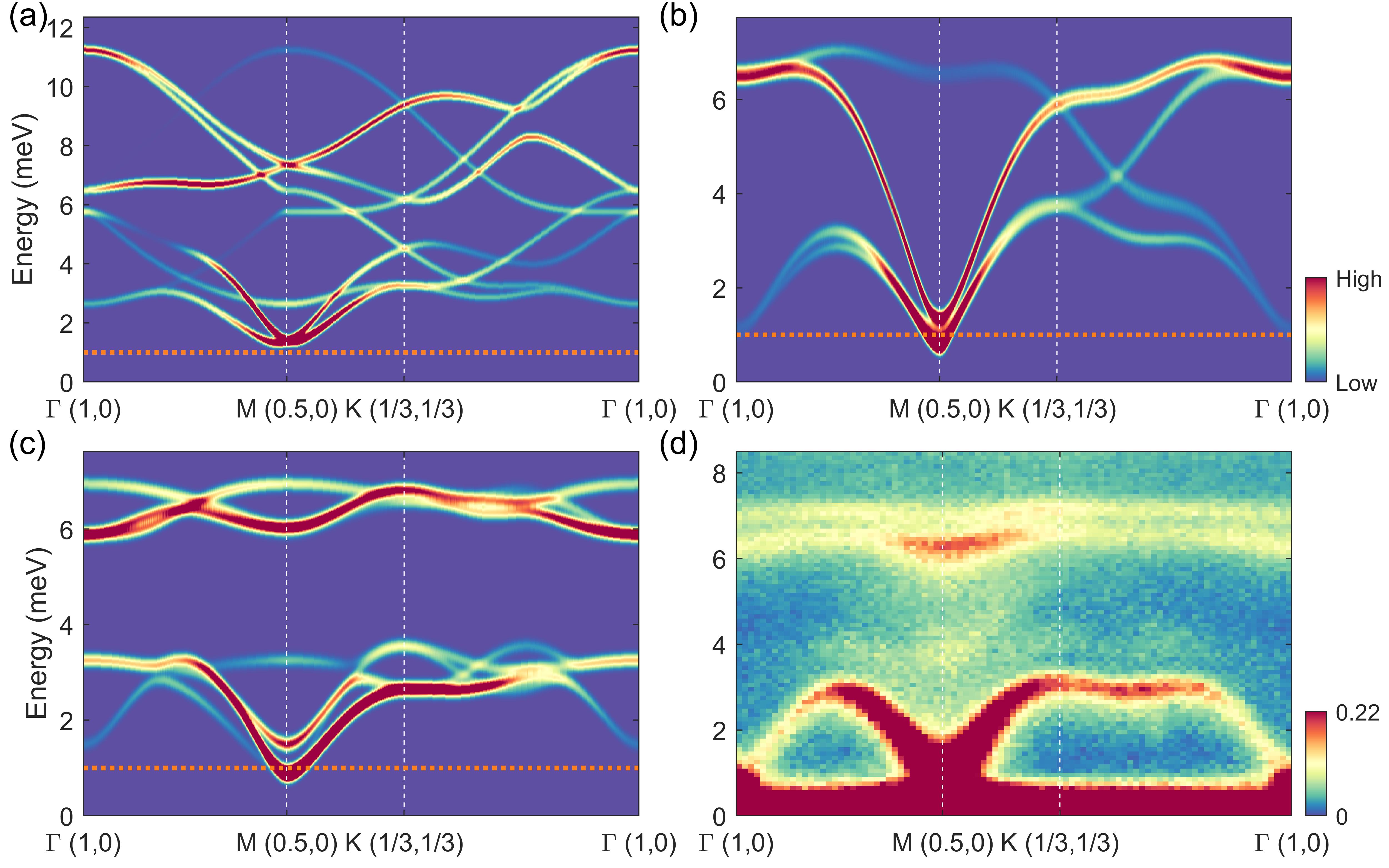}}
		\caption{(a)-(c) Calculated spin wave spectra with the best-fit parameters reported in \cite{SongvilayPRB2020,LinNC2021,KimJPCM2021}, which are displayed in the same trajectory as in Fig. 3(a) of the main text. Horizontal dashed lines indicate 1 meV. (d) Same data as the lower part in Fig. 3(a) of the main text.}
	\label{figs4}
	\end{figure}

	We calculated the spin wave spectra with three sets of parameters reported in \cite{SongvilayPRB2020,LinNC2021,KimJPCM2021}(see Table \ref{tb1}). The results are presented in Fig. \ref{figs4}(a)-(c). We find none of these models can reproduce the measured excitation spectrum in Fig. \ref{figs4}(d). However, the model proposed by Lin \textit{et al.} most correctly accounts for close gap sizes ($\sim$1 meV) at $\Gamma$-point and M-point. Their model has a dominant $J_3$ term, which is qualitatively in line with our finding.
	
	\begin{table}[b]
		\caption{\label{tab:table1}
			Best-fit parameters of $H$-$K$-$\Gamma$ model reported by recent three representative publications about \ch{Na_2Co_2TeO_6}.
		}
		\begin{ruledtabular}
			\begin{tabular}{ccccccc}
				Interactions (in meV)&$J_1$&$J_2$&$J_3$&$K$&$\Gamma$&$\Gamma^\prime$\\
				\midrule
				Songvilay \textit{et al.} \cite{SongvilayPRB2020}&-0.1(5)&0.3(3)&0.9(3)&-9.0(5)&1.8(5)&0.3(3)\\Lin \textit{et al.} \cite{LinNC2021}
				&-2.325&0&2.5&0.125&0.125&0\\Kim \textit{et al.} \cite{KimJPCM2021}
				&-1.50(5)&0&1.50(2)&3.30(10)&-2.80(5)&2.10(7)\\
			\end{tabular}	
		\end{ruledtabular}
		\label{tb1}
	\end{table}

	\section{Powder averaged single crystal data}
	
\begin{figure}[h!]
		\centering{\includegraphics[clip,width=14cm]{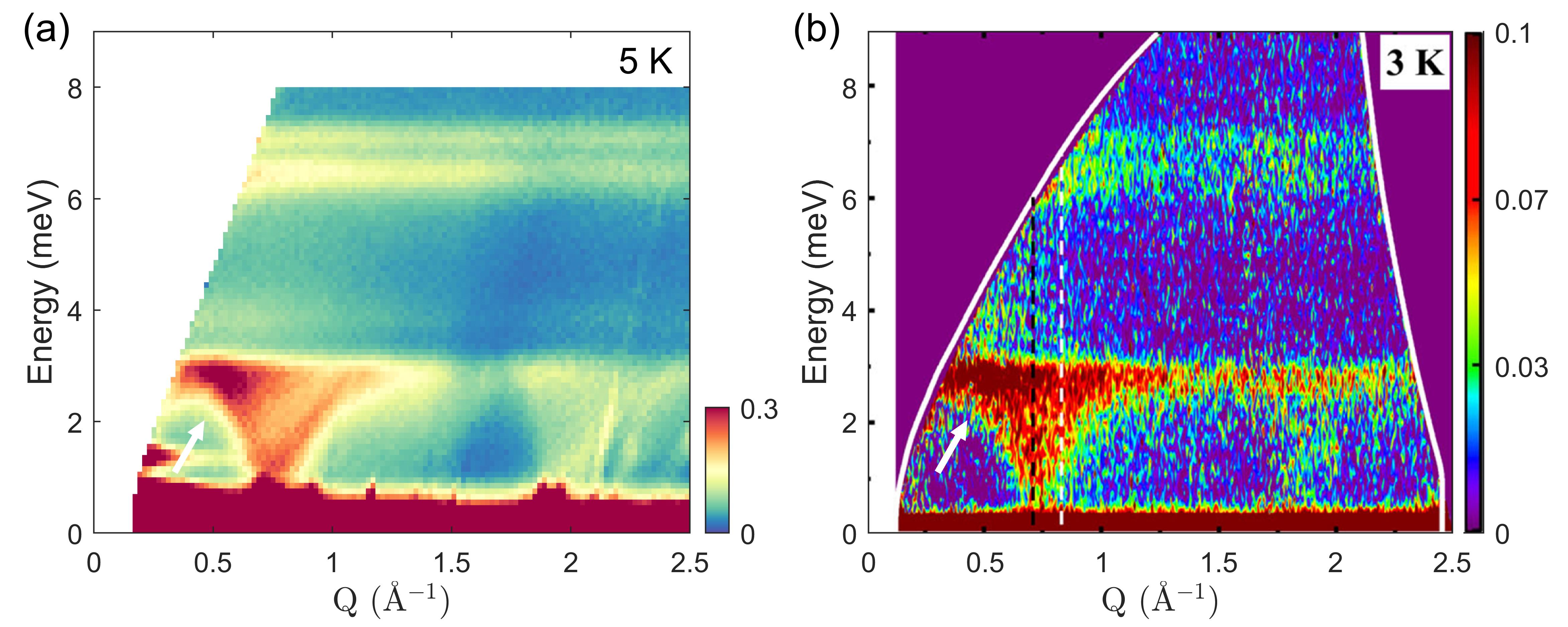}}
		\caption{(a) Powder averaged INS spectrum made from our single crystal data at 5 K with $E_\mathrm{i}$ = 10.0 meV. (b) Real powder INS spectrum adapted from \cite{LinNC2021}. The white arrows in both panels indicate the concave shape.}
		\label{figs5}
	\end{figure}

	We made a powder averaged plot for our single crystal INS data at 5 K with $E_\mathrm{i}$ = 10.0 meV, which is presented in Fig. \ref{figs5}(a). For comparison, one powder INS spectrum adapted from \cite{LinNC2021} is presented in Fig. \ref{figs5}(b). The intensity in high-$Q$ part of Fig. \ref{figs5} (a) may suffer from the anisotropy of our single crystal data. We note all spin wave modes identified in powder INS experiments are reproduced, which confirms that our INS data are fully consistent with previous reports \cite{SongvilayPRB2020,LinNC2021,SamarakoonPRB2021,KimJPCM2021,SandersArxiv2021}. In particularly, we can clearly see the concave shape close to the magnitude of the M-point, which has been generally regarded as an evidence for the underlying zigzag order.

	\section{Additional quasielastic scattering data with Ei = 19.4 meV}
	
\begin{figure}[h]
		\centering{\includegraphics[clip,width=12cm]{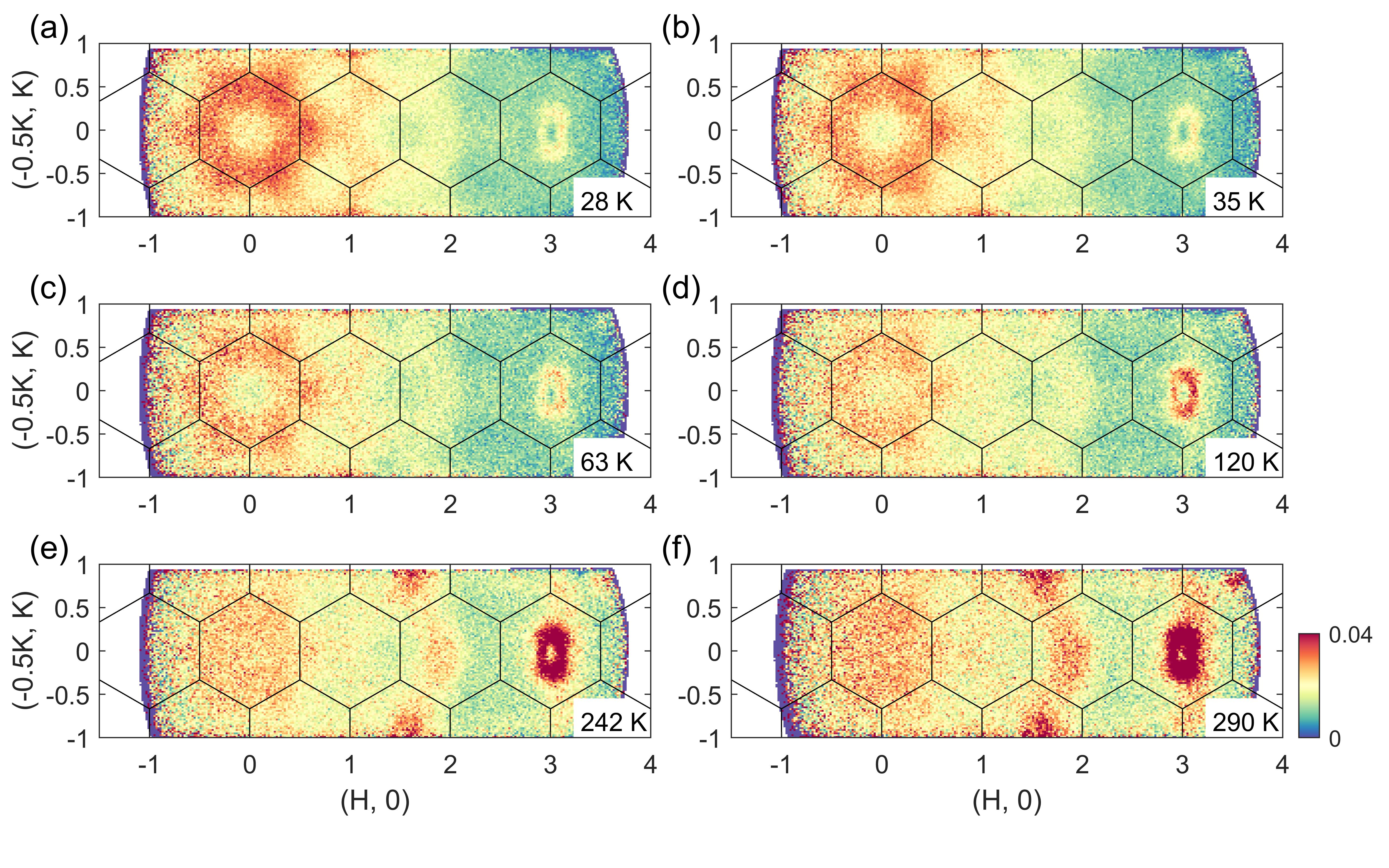}}
		\caption{Constant energy cuts in the ($H$, $K$) plane at six temperatures above $T_N$. Energy is integrated from 5 meV to 9 meV. Data are measured with $E_\mathrm{i}$ = 19.4 meV. Phonon background correction was not made.}
	\label{figs6}
	\end{figure}

	Intensity distributions above $T_\mathrm{N}$ in a wider range of ($H$, $K$) plane are presented in Fig. \ref{figs6}. The quasielastic scattering feature is consistent with equal-time spin correlations based on zigzag-typed spin arrangements as discussed in the main text. Phonon scattering around (3, 0) becomes obvious as the temperature goes up. The intensity close to $H$ = 2 comes from the aluminum sample holder.
	
	\section{Other calculated spin correlation patterns}

\begin{figure}[h!]
		\centering{\includegraphics[clip,width=10cm]{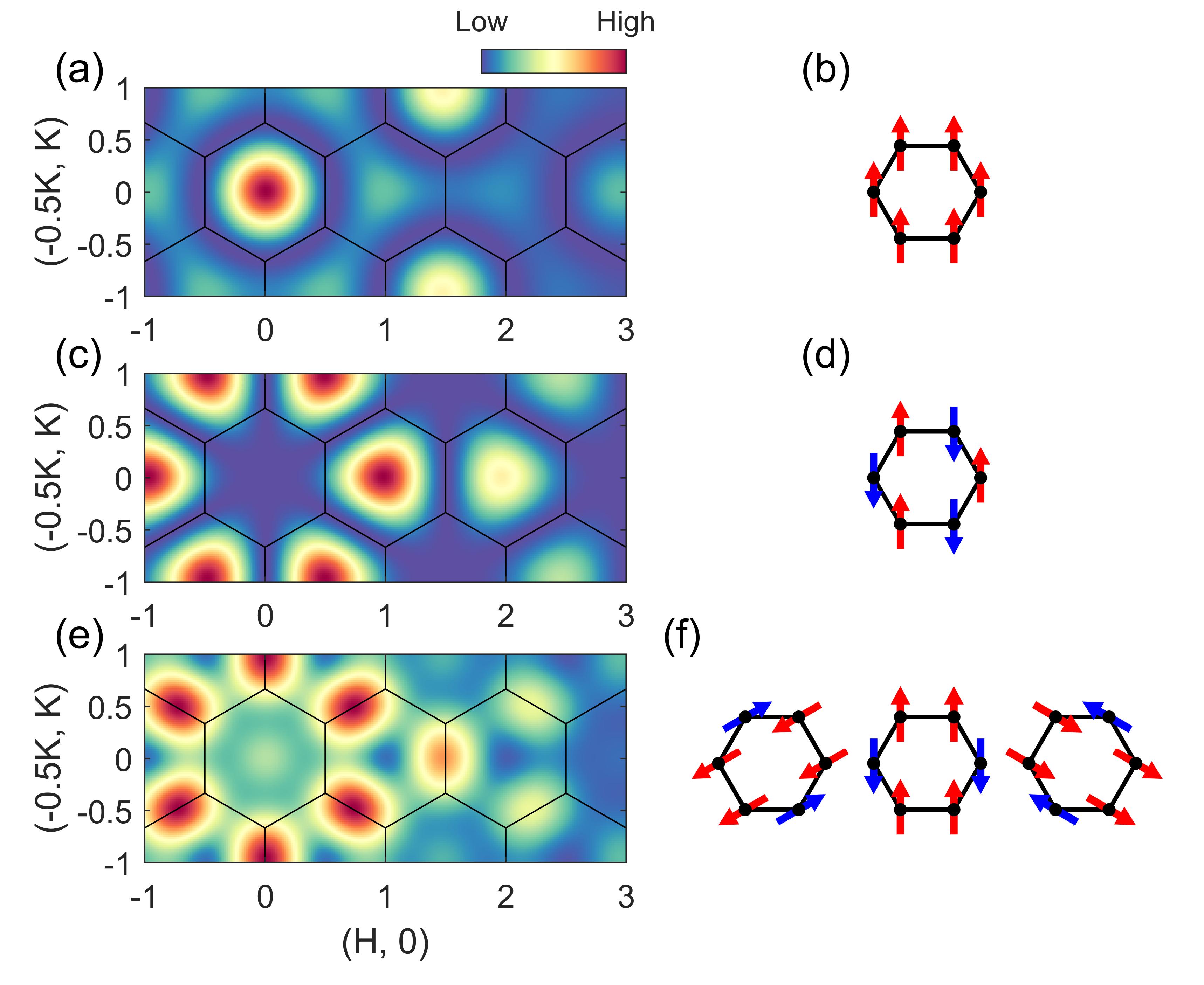}}
		\caption{(a), (c) and (e) Calculated structure factors in the ($H$, $K$) plane for ferromagnetic, N\'eel and stripy spin arrangements presented in (b), (d) and (f), respectively.}
		\label{figs7}
	\end{figure}

	Fig. \ref{figs7} presents three calculated equal-time spin correlation patterns based on ferromagnetic, N\'eel and stripy arrangements within one hexagonal unit, which are inconsistent with our experimental observation.
	
\begin{figure}[h!]
		\centering{\includegraphics[clip,width=12cm]{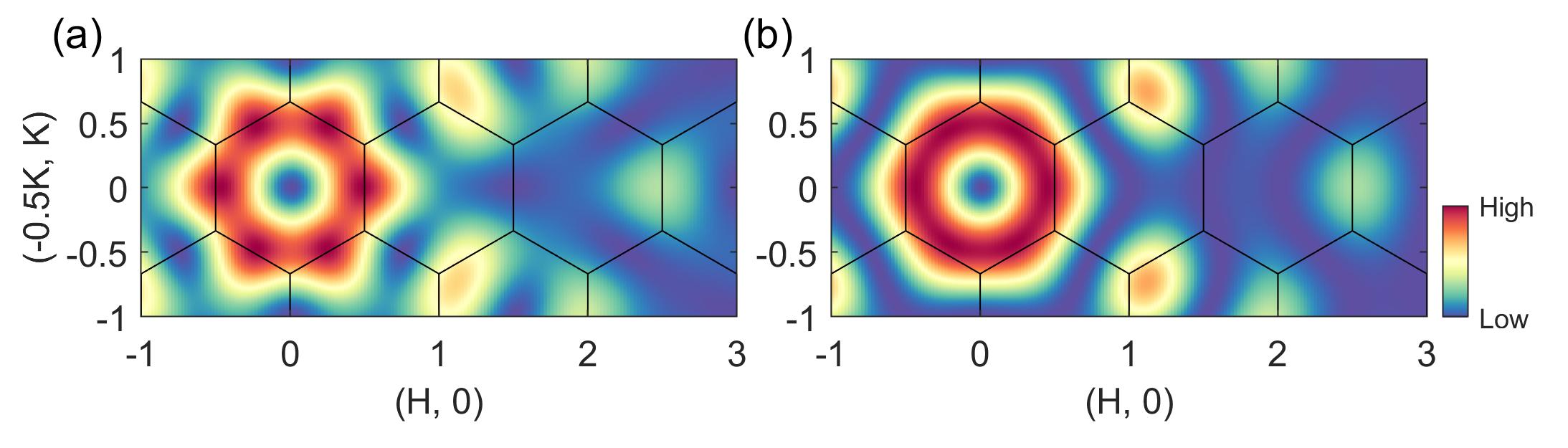}}
		\caption{(a) and (b) Calculated static spin correlation patterns in the ($H$, $K$) plane for the collinear [Fig. 4(e) left] and non-collinear [Fig. 4(e) right] zigzag-typed spin arrangements, respectively.}
		\label{figs8}
	\end{figure}

	To see the fluctuating nature of the spins in the paramagnetic state, we calculated the static spin correlation patterns for the zigzag and superposed spin arrangements within one hexagonal unit [Fig. 4(e) in the main text]. The intensity can be written as
	\begin{equation}
		I(\textbf{Q}) = f^2(\textbf{Q}) \sum_{m,n}e^{i\textbf{Q}\cdot(\textbf{r}_m-\textbf{r}_n)}\sum_{\alpha,\beta}(\delta_{\alpha,\beta}-\frac{Q_{\alpha}Q_{\beta}}{Q^2}) S_m^{\alpha}S_n^{\beta},
	\end{equation}
	where the 4$\pi$ rotation average is not performed comparing with (2) in the main text. The calculated results can be found in Fig. \ref{figs8}. Although their main features are similar with the equal-time spin correlation patterns, the latter are more consistent with the observed pattern in detail, reflecting the fact that the spins are indeed fluctuating in the paramagnetic state.

\end{document}